\documentclass[runningheads]{llncs}
\usepackage[T1]{fontenc}
\usepackage{amsfonts}       
\usepackage{amsmath}
\usepackage{bm}
\usepackage{bbm}
\usepackage[numbers, square, sort&compress]{natbib}
\usepackage{hyperref}
\usepackage{graphicx}
\usepackage{algorithm,algpseudocode}

\algnewcommand{\Inputs}[1]{%
  \State \textbf{Inputs:}
  \Statex \hspace*{\algorithmicindent}\parbox[t]{\linewidth}{\raggedright #1}
}
\algnewcommand{\Output}[1]{%
  \State \textbf{Output:}
  \Statex \hspace*{\algorithmicindent}\parbox[t]{\linewidth}{\raggedright #1}
}
\algnewcommand{\Initialize}[1]{%
  \State \textbf{Initialize:}
  \Statex \hspace*{\algorithmicindent}\parbox[t]{\linewidth}{\raggedright #1}
}
\algnewcommand{\comm}[1]{ {\ttfamily\textcolor{blue}{// #1}} }
\usepackage{subfigure}
\usepackage{tabularx}
\usepackage{multirow}
\usepackage{booktabs}
\usepackage{url}

\usepackage{xcolor}

\graphicspath{ {./figures/} }

\begin{document}

\title{Achilles' Heels: Vulnerable Record Identification in Synthetic Data Publishing}

\titlerunning{Achilles' Heels}

\author{Matthieu Meeus\textsuperscript{*}  \and
Florent Gu\'{e}pin\textsuperscript{*}  \and
Ana-Maria Cre\c tu \and
Yves-Alexandre de Montjoye 
}

\authorrunning{M. Meeus, F. Gu\'{e}pin et al.}

\institute{Department of Computing and Data Science Institute, 
Imperial College London, London, United Kingdom \\
\email{\{m.meeus22,florent.guepin20,a.cretu,deMontjoye\}@imperial.ac.uk}}

\maketitle              
\def\thefootnote{*}\footnotetext[1]{These authors contributed equally to this work.}

\begin{abstract}

Synthetic data is seen as the most promising solution to share individual-level data while preserving privacy. Shadow modeling-based Membership Inference Attacks (MIAs) have become the standard approach to evaluate the privacy risk of synthetic data. While very effective, they require a large number of datasets to be created and models trained to evaluate the risk posed by a single record. The privacy risk of a dataset is thus currently evaluated by running MIAs on a handful of records selected using ad-hoc methods. We here propose what is, to the best of our knowledge, the first principled vulnerable record identification technique for synthetic data publishing, leveraging the distance to a record's closest neighbors. We show our method to strongly outperform previous ad-hoc methods across datasets and generators. We also show evidence of our method to be robust to the choice of MIA and to specific choice of parameters. Finally, we show it to accurately identify vulnerable records when synthetic data generators are made differentially private. The choice of vulnerable records is as important as more accurate MIAs when evaluating the privacy of synthetic data releases, including from a legal perspective. We here propose a simple yet highly effective method to do so. We hope our method will enable practitioners to better estimate the risk posed by synthetic data publishing and researchers to fairly compare ever improving MIAs on synthetic data. 

\end{abstract}

\section{Introduction}
\label{sec:introduction}

There is increased demand from businesses, governments, and researchers to make data widely available to support research and innovation~\cite{fca22synthetic}, including the development of Artificial Intelligence (AI) models.
Data, however, most often relates directly or indirectly to individuals, raising privacy concerns.

Synthetic data is seen as a promising solution to share individual-level data while preserving privacy \cite{bellovin2019privacy}.
Synthetic data is generated by sampling new data values from a statistical model whose parameters are computed from the original, private data. A large range of techniques have been proposed to generate synthetic data \cite{zhang2017, nowok2016synthpop, xu2019modeling}. 
Synthetic data, if truly privacy-preserving, can be shared and used freely as it no longer falls under the scope of data protection legislation such as the European Union's General Data Protection Regulation (EU GDPR) or California's CCPA. 
Consequently, synthetic data has garnered significant interest from statistical offices~\cite{bates2019ons}, health care agencies~\cite{nhs2020aande} and the financial sector~\cite{fca22synthetic}. 

Ensuring that synthetic data preserves privacy is, however, difficult. On the one hand, while the statistical model aggregates the original data, it is well known that aggregation alone is not sufficient to preserve privacy~\cite{dinur2003revealing, pyrgelis2017knock}. On the other hand, ensuring that synthetic data generation models achieve formal privacy guarantees is difficult and comes at a cost in utility \cite{annamalai2023linear, stadler2022synthetic}. 

Membership Inference Attacks (MIA) are a key tool to evaluate the privacy guarantees offered by synthetic data generation models, differentially private or not, in practice \cite{houssiau2022tapas}. If successful, an MIA is able to infer that a particular record was part of the original data used to train the synthetic data generation model. Increasingly advanced MIAs have been proposed against synthetic data \cite{stadler2022synthetic, houssiau2022tapas}. They are run on a per-record basis, instantiating an attacker that aims to distinguish between two worlds: (1) a world in which the synthetic dataset released was generated from a model fitted on data \textit{including} the target record and (2) the alternative world in which the model was fitted on data \textit{excluding} the target record. MIAs on synthetic data have already successfully disproved the believe that aggregation alone is sufficient to preserve privacy as well as detected issues with differentially private synthetic data generation models \cite{stadler2022synthetic}.

Fully evaluating the privacy risk of a dataset using MIAs is however out-of-reach. Indeed state-of-the-art MIAs rely on shadow models which require training a large number, often in the order of thousands, of synthetic data generators to learn the optimal decision boundary for each record. Running a state-of-the-art MIA on a record takes in our experiments, depending on the generator and the dataset, approximately 1.5 to 6 hours on dedicated computation facilities. Even for the (relatively) small datasets of 1,000 records we consider, fully estimating the privacy risk of a dataset, would take up to 250 days of compute. 

Current evaluation of the privacy risk posed by synthetic data, and of their privacy-utility trade-off, is thus currently the result of ever more advanced MIAs evaluated on only a handful of records selected using ad-hoc methods. 
Stadler et al.~\cite{stadler2022synthetic} evaluate their attack on ten records, half random and half hand-picked outliers selected based on a rare attribute value, while Houssiau et al.~\cite{houssiau2022tapas} evaluate their attack on one record selected based on the lowest log-likelihood.

Such ad-hoc approaches to selecting vulnerable records could strongly underestimate the privacy risks of a synthetic data release, missing highly vulnerable records or wrongly concluding that most records are not at risk. 

\textbf{Contribution.} We here propose what is, to the best of our knowledge, the first vulnerable record identification strategy for synthetic data. 

We formalize the problem and propose a principled and simple metric to identify vulnerable records in the original, private dataset: the mean distance to its closest neighbors.
While simple, we show our method to outperform previous ad-hoc approaches by 7.2 percentage points (p.p.) on average on two datasets and two generators. Both when comparing the performance of ever improving attack methodologies as well as to estimate the potential privacy risk in real world synthetic data publishing, this is a significant difference. 

We then extensively evaluate our method. First, we evaluate its applicability across different MIA approaches.
We develop a fundamentally different MIA approach and show the risk estimated using our vulnerable record identification method to be consistently and significantly higher compared to prior methods.
More specifically, in contrast with Houssiau et al.~\cite{houssiau2022tapas}, who train a meta-classifier to infer the membership of a record based on hand-engineered aggregate statistics computed on the synthetic dataset, we develop an attack that trains a meta-classifier directly on the synthetic dataset viewed as \textit{a set of records}. After performing the same extensive empirical evaluation, we find that the performance of our attack increases also significantly when computed on records identified by our method, by 5.2 p.p. on average across datasets and generators. 
Next, we evaluate the sensitivity of our results on both the number of neighbors included in our metric as well as the distance metric. In both cases, we find that the results do not change significantly and confirm that our chosen metric is justified. 
Finally, we evaluate our metric on a differentially private synthetic data generator for varying values of the privacy budget $\epsilon$. We confirm that MIAs fail when the privacy budget is decreased to $\epsilon=1$, and find that our metric consistently identifies more vulnerable records for larger values of $\epsilon$. 

With this principled method, we hope to eliminate the need to rely on ad-hoc record selection and potentially underestimate the privacy risk in synthetic data publishing. A formal vulnerable record identification method would enable (1) researchers to fairly compare future state-of-the-art attacks and (2) practitioners to, in line with EU legislations, evaluate the privacy protection of synthetic data on the worst-case scenario.

\section{Background}
\label{sec:background}

\subsection{Synthetic Data Generation}
\label{subsec:background:synthetic_data}
We consider an entity (e.g., a company) that wants to give a third-party access to a private dataset for analysis.
A dataset is a multiset of records $D=\{ x_1, \ldots, x_n \}$, where each record relates to a unique individual. In the case of narrow datasets, multiple individuals can however have the same record. We assume the dataset to be sensitive, containing information such as healthcare records or financial transactions. 
We assume each record to consist of $F$ attributes $x_i=(x_{i,1}, \ldots, x_{i,F}) \in \mathcal{V}_1 \times \ldots \times \mathcal{V}_F$, where $\mathcal{V}_j$ denotes the space of values that can be taken by the $j$-th attribute.
We denote by $\mathcal{V}=\mathcal{V}_1 \times \ldots \times \mathcal{V}_F$ the universe of possible records, and assume the dataset to be sampled i.i.d. from a distribution $\mathcal{D}$ over $\mathcal{V}$.

An increasingly popular approach to release data while mitigating the 
privacy risk is to instead generate and publish a synthetic dataset \cite{edge2020design}.
Synthetic data is generated from a statistical model trained to have similar statistical properties to the real data.

We refer to the statistical model as the \textit{synthetic data generator}.
Formally, this is a randomized function $\Phi: \mathcal{V}^n \rightarrow \mathcal{V}^{m}$ mapping a private dataset $D$ to a synthetic dataset $D^s = \Phi(D)$ of $m$ records. 
The synthetic data generator can take the form of a probabilistic model such as a Bayesian Network \cite{zhang2017} or a Generative Adversarial Network (GAN) such as CTGAN \cite{ctgan2019}.

\subsection{Differential Privacy}
\label{subsec:background:differential_privacy}

Differential Privacy (DP) \cite{dwork2006differential} is a formal privacy guarantee. Originally developed to protect the release of aggregate statistics, it has since been extended to machine learning models and recently to synthetic data generators.

DP relies on a notion of \textit{neighboring} datasets. Two datasets $D, D' \sim \mathcal{D}$ are neighboring, if they differ by only one record. Intuitively, a DP randomized function $\Phi$ ensures that one attacker cannot infer the difference between $\Phi(D)$ and $\Phi(D')$ bounded by a certain probability, effectively providing formal guarantees against privacy risks and, in particular, membership inference attacks. 

Formally, the definition of differential privacy is the following: 
\begin{definition}[$\epsilon$\textbf{-Differential Privacy}]
    A randomised function $\Phi$ is said to be $\epsilon$-Differentially Private ($\epsilon$-DP) if for every pair of neighboring datasets $D,D'$, and for all subsets of outputs $S \in Range(\Phi)$, 
    the following inequality holds:
    $\mathbb{P}[\Phi(D) \in S] \leq e^{\epsilon}\mathbb{P}[\Phi(D') \in S]$.
\end{definition}

The parameter $\epsilon$ is referred to as the privacy budget.
It quantifies the privacy leakage of a data release, with smaller $\epsilon$ providing more protection. 
Achieving DP with small to reasonable values of $\epsilon$ can however require significant noise to be added to a machine learning model, decreasing its utility. 
Efficiently achieving DP guarantees for ML models, including synthetic data generation models, is an active area of research. In this paper, we used the DP generator PrivBayes from the work of Zhang et al. \cite{zhang2017}. We refer the reader to Sec. \ref{subsec:privBayes} for details. 

\subsection{Membership Inference Attacks}
\label{subsec:background:mia}
\textbf{Threat Model.} 
Membership Inference Attacks (MIAs) aim to infer whether a target record $x_T$ is part of the original dataset $D$ used to generate a synthetic dataset $D^s$, i.e., whether $x_T \in D$.
The traditional attacker in MIAs is assumed to have access to the published synthetic dataset, $D^s$, as well as to an auxiliary dataset, $D_{aux}$, with the same underlying distribution as $D$: $D_{aux} \sim \mathcal{D}$. Additionally, the attacker is assumed to have access to the target record $x_T$ and to know the generative model type and its hyperparameters, but not to have access to the model \cite{stadler2022synthetic,houssiau2022tapas}. 

\textbf{Privacy Game.}
MIAs are instantiated as a privacy game. 
The game consists of a challenger and an attacker and is instantiated on a target record $x_T$.
The challenger samples datasets $D$ of $n-1$ records from $\mathcal{D}$, such that all records are different from $x_T$. 
With equal probability, the challenger adds $x_T$ to $D$ or a random, different record $x_R \sim \mathcal{D}$, with $x_R \neq x_T$, to $D$.
The challenger then trains a generator $\Phi_T$ on $D$ and uses it to generate a synthetic dataset $D^s$.
The challenger shares $D^s$ with an attacker whose goal is to infer whether or not $x_T \in D$.
If the attacker correctly infers, they win the game.
The privacy risk is estimated by playing the game multiple times and reporting the average.

\textbf{Shadow Modeling.}
State-of-the-art MIAs against synthetic data generators rely on the shadow modeling technique \cite{shokri2017membership,stadler2022synthetic,houssiau2022tapas}. 
With this technique, the attacker leverages their access to the auxiliary dataset as well as the knowledge of the model, to train multiple instances of $\Phi_T$ (called $\Phi_{shadow}$) and evaluate the impact of the presence or absence of a target record on the resulting model. 
An attacker aiming to perform an MIA will proceed as follows. They will first create (e.g. by sampling) multiple datasets $D_{shadow}$ from $D_{aux}$, such that $|D_{shadow}| = |D|-1$. Note that we ensure $x_T \notin D_{aux}$. The attacker then adds $x_T$ to half of the $D_{shadow}$ and a random record $x_R$, distinct from $x_T$, to the other half. 
Using the completed $D_{shadow}$ (now $|D_{shadow}|=|D|$) and the knowledge of model $\Phi_T$, the attacker will now train multiple shadow generators $\Phi_{shadow}$. 
The attacker will then use the $\Phi_{shadow}$ to produce synthetic datasets $D_{shadow}^s$, each labeled with either \textit{IN} if $x_T \in D_{shadow}$ or $\textit{OUT}$ otherwise. 
Using the labeled dataset, the attacker can now train a meta-classifier $\mathcal{M}_{meta}$ to distinguish between cases where the target record was and was not part of the training dataset. More specifically, $\mathcal{M}_{meta}$ would be trained using $D_{shadow}^s$ and the constructed binary label \textit{IN} or \textit{OUT}. At inference time, the attacker would then query the $\mathcal{M}_{meta}$ on the released $D^s$ for records of interest, and return its prediction.

\section{Related Works}
\label{sec:related_works}

\textbf{Membership Inference Attacks (MIAs)} were first developed by Homer et al.~\cite{homer2008resolving} to study whether the contribution of an individual to a Genome-Wide Associated Study can be inferred from released genomic aggregates.
Their attack was based on a statistical test aiming to distinguish between aggregates that include the individual and aggregates computed on individuals randomly drawn from a reference population.
Sankararaman et al.~\cite{sankararaman2009genomic} extended the analysis soon after and showed the risk of membership inference to increase with the number of aggregates released $m$, but decrease with the number of individuals $n$ in the dataset.
MIAs have since been widely used to evaluate the privacy risk in aggregate data releases such as location~\cite{pyrgelis2017knock} or survey data~\cite{bauer2020towards}.

\textbf{MIAs Against Machine Learning (ML) Models.} 
Shokri et al.~\cite{shokri2017membership} proposed the first MIA against ML models. Their attack relies on a black-box access to the model and the shadow modeling technique (see Sec.~\ref{subsec:background:mia} in the context of synthetic data generators).
MIAs against ML models have since been extensively studied in subsequent works~\cite{yeom2018privacy,salem2018ml,truex2019demystifying,nasr2019comprehensive,leino2020stolen}, both from a privacy perspective but also to better understand what a model learns. These works have, for instance, shown the risk to be higher in overfitted models~\cite{yeom2018privacy}  
and smaller datasets~\cite{shokri2017membership}, and to be mitigated by differentially private training, albeit at a cost in utility~\cite{jayaraman2019evaluating,leino2020stolen}. 

\textbf{Disparate Vulnerability of Records.}
Previous work on MIA against ML models has shown that not all records are equally vulnerable to MIAs.
The measured risk has e.g. been shown to vary with the label~\cite{shokri2017membership} and to be higher for outlier records~\cite{long2020pragmatic,carlini2022privacy} and members of subpopulations~\cite{chang2021privacy}. 
Feldman proposed a theoretical model demonstrating that models may, in fact, need to memorise rare or atypical examples in order to perform optimally when trained on long-tailed distributions such as those found in modern datasets~\cite{feldman2020does,feldman2020neural}.
Carlini et al.~\cite{carlini2022privacy} showed this effect to be relative, as removing the most vulnerable records increases the vulnerability of remaining ones. 
Importantly for this work, Long et al. \cite{long2020pragmatic} argued that attacks should be considered a risk even when they only perform well on specific records.
Long et al. proposed an approach to select vulnerable records prior to running the attack so as to increase its efficiency and precision. While their work \cite{long2020pragmatic} shows that records further away from their neighbors are more vulnerable, it only considers ML classification models.

\raggedbottom
\textbf{MIAs Against Synthetic Tabular Data.} 
MIAs have been extended to synthetic tabular data, the focus of our work. 
From a methodological standpoint, they fall broadly into two classes.
The first class of methods directly compares synthetic data records to the original, private records, searching for exact or near-matches \cite{domingo2015disclosure,lu2019empirical,yale2019assessing,giomi2022unified}.
The second class of methods instantiates the shadow modeling technique in the black-box setting, assuming the adversary has knowledge of the algorithm used to generate synthetic data and to an auxiliary dataset. 
Stadler et al.~\cite{stadler2022synthetic} trained a meta-classifier on aggregate statistics computed on the synthetic shadow datasets, specifically the mean and standard deviation of the attributes, correlation matrices and histograms.
Houssiau et al.~\cite{houssiau2022tapas} extended this work using $k$-way marginal statistics computed over subsets of the attribute values of the targeted record. 
They also extended the threat model from the standard black-box setting, which assumes knowledge of the generative model used, to the no-box setting that lacks this assumption.
In this paper, we focus exclusively on shadow modeling-based attacks which are the state of the art.

\section{Identifying Vulnerable Records}
\label{sec:methods}

We here propose and validate a simple, yet effective, approach to identify vulnerable records in synthetic data publishing.
Our approach is motivated by findings of previous works in the ML literature that records further away from their neighbors present a higher risk compared to randomly selected records.
To identify such records, we compute a distance metric $d(x_i, x_j)$ between every pair of records in the dataset $(x_i, x_j) \in D \times D$, with $i \neq j$\footnote{Note that since the dataset $D$ can contain repeated records (i.e., two or more individuals sharing the same attributes), the closest record to $x_i$ can be a duplicate $x_j=x_i, j \neq i$ such that the distance between them is zero: $d(x_i, x_j)=0$.}.
Then, for every record $x_i \in D$, we define its \textit{vulnerability score} $V_k(x_i)$ as the average distance to its closest $k$ neighbors in the dataset.

\begin{definition}[Vulnerability Score]
Given a dataset $D$, a record $x_i \in D$, and a distance metric $d$, the vulnerability score of the record is defined as $V_k(x_i)= \frac{1}{k}\sum_{j=1}^k d(x_i, x_{i_j})$, where $x_{i_1}, \ldots, x_{i_{|D|-1}}$ is the re-ordering of the other records according to their increasing distance to $x_i$, i.e., $d(x_i, x_{i_1}) \leq \cdots \leq d(x_i, x_{i_{|D|-1}})$, with $i_j \neq i$ for every $j =1, \ldots, |D|-1$.
\end{definition}

Finally, we rank the records in $D$ decreasingly by their vulnerability score $V_k(x_{r_1}) \geq \ldots \geq V_k(x_{r_{|D|}})$ and return the top-$R$ records as the most vulnerable records. In the unlikely event that the last record(s) included in the top-$R$ and the first one(s) not included have the exact same value for the vulnerability score, then a random subset is chosen.

Our distance metric $d$ carefully distinguishes between categorical and continuous attributes.
Recall that the dataset consists of $F$ attributes.
When the space of values that can be taken by an attribute is discrete, e.g., the gender of an individual $\mathcal{V}_f=\{ \text{female}, \text{male} \}$, we refer to the attribute as \textit{categorical}, while when the space of values is continuous, e.g., the income of an individual, we refer to the attribute as \textit{continuous}.
Denote by $\mathcal{F}_{\text{cont}}$ and $\mathcal{F}_{\text{cat}}$ the subsets of continuous and categorical attributes, with $\mathcal{F}_{\text{cont}} \cup \mathcal{F}_{\text{cat}} = \{1, \ldots, F \}$ and $\mathcal{F}_{\text{cont}} \cap \mathcal{F}_{\text{cat}} = \emptyset$.
We preprocess each record $x_i = (x_{i,1}, \ldots, x_{i,F}) \in D$ as follows:

\begin{enumerate}
    \item We convert every categorical attribute $x_{i,f}$, with $f \in \mathcal{F}_{\text{cat}}$ and $|\mathcal{V}_f|$ possible values $v_1, \ldots, v_{|\mathcal{V}_f|}$ to a one-hot encoded vector: \begin{equation}
h(x_{i,f}) := (\mathbbm{1}(x_{i,f}=v_1), \ldots, \mathbbm{1}(x_{i,f}=v_{|\mathcal{V}_f|})),
\end{equation}
i.e., a binary vector having 1 in the $q$-th position if and only if the attribute value is equal to $v_q$. 
Here, $\mathbbm{1}$ denotes the indicator function.
We then concatenate the one-hot encoded categorical attributes into a single vector:
\begin{equation}
    h(x_i) := (h(x_{i,f}))_{f \in \mathcal{F}_{\text{cat}}}
\end{equation}

\item We normalise every continuous attribute $x_{i,f}$, with $f \in \mathcal{F}_{\text{cont}}$ to range between 0 and 1, using min-max scaling:
\begin{equation}n(x_{i,f}) := \frac{x_{i,f} - m_f(D)}{ M_f(D) - m_f(D)},
\end{equation}
that is, we scale the values using the minimum $m_f(D) := \min\limits_{j=1,\ldots,|D|}(x_{j,f})$ and maximum  $M_f(D) := \max\limits_{j=1,\ldots,|D|}(x_{j,f})$ values of the attribute estimated from the dataset $D$.
We then concatenate the normalised continuous attributes into a single vector:
\begin{equation}
    c(x_i) := (n(x_{i,f}))_{f \in \mathcal{F}_{\text{cont}}}
\end{equation}

\end{enumerate}

Using this notation, we define our distance metric between two records $x_i$ and $x_j$ as follows:

\begin{equation}
    d(x_i, x_j) := 1 
    - \frac{|\mathcal{F}_{\text{cat}}|}{F} \underbrace{
    \frac{h(x_i) \boldsymbol{\cdot} h(x_j)}{||h(x_i)||_2 * ||h(x_j)||_2}
    }_{\substack{\text{cosine similarity between} \\ \text{categorical attributes}}}
    - \frac{|\mathcal{F}_{\text{cont}}|}{F} 
    \underbrace{\frac{c(x_i) \boldsymbol{\cdot} c(x_j)}{||c(x_i)||_2*||c(x_j)||_2}}_{\substack{\text{cosine similarity between} \\ \text{continuous attributes}}},
\label{eq:distance}
\end{equation}
where $\boldsymbol{\cdot}$ denotes the dot product between two vectors, $*$ denotes scalar multiplication, and $||a||_2 := \sqrt{\sum_{l=1}^{L} a_l^2}$ is the Euclidean norm of a vector $a \in \mathbb{R}^L$.

The resulting distance is a generalisation of the cosine distance across attribute types.
It ranges between 0 and 1, with larger values indicating records to be less similar. 
The distance is equal to 0 if and only if the records are identical.

\section{Experimental Setup}
\label{sec:experimental_setup}

\subsection{State-of-the-art MIA Against Synthetic Tabular Data}
\label{subsec:SOTA_MIA}
Houssiau et al.~\cite{houssiau2022tapas} developed an MIA that instantiates the shadow modeling technique (see Sec.~\ref{subsec:background:mia} for details) by training a random forest classifier on a subset of $k$-way marginal statistics that select the targeted record.
The authors evaluated this approach on only one outlier record and found it to outperform other methods.
Throughout our experiments, we found the attack to consistently outperform alternative methods and thus confirm it to be the state-of-the-art attack against synthetic tabular data. We refer to it as the \textit{query-based attack}. 

Given a target record $x_T$ and a synthetic dataset $D^s$ sampled from a generator fitted on a dataset $D$, 
this attack computes \emph{counting queries} $Q^A$ to determine how many records in the synthetic dataset match a subset $A\subset \{1, \ldots, F \}$ of attribute values of target record $x_T$. We denote by $a_i$ the $i$-th attribute for every $x\in D^s$ in the following:
$
Q^A(x_T) = \text{{COUNT}}\text{ WHERE } \bigwedge_{i \in A} (a_i = x_{T,i}) 
$.
Counting queries are equal to $k$-way marginal statistics computed on the attribute values of the target record,  multiplied by the dataset size $|D^s|$. 

Our intuition behind this attack is twofold.
First, when these statistics are preserved between the original and synthetic datasets, the answers are larger by 1, on average, when $x_T \in D$ compared to when $x_T \notin D$, since the queries select the target record.
Second, the impact of the target record on the synthetic dataset is likely to be local, i.e., more records matching a subset of its attribute values are likely to be generated when $x_T \in D$, leading to statistically different answers to the same queries depending on the membership of $x_T$.

Like Houssiau et al. \cite{houssiau2022tapas}, we randomly sample a predetermined number $N$ of attribute subsets $A$ and feed the answers to the associated $Q^A$, computed on the synthetic shadow dataset, to a random forest meta-classifier.
For computational efficiency reasons, we here use a C-based implementation of the query computation \cite{cretu2022querysnout} instead of the authors' Python-based implementation~\cite{houssiau2022tapas}.

We further extend the method to also account for continuous columns by considering the less-than-or-equal-to $\leq$ operator on top of the equal operator $=$ that was previously considered exclusively. For each categorical attribute we then use $=$, while using the $\leq$ for each continuous attribute.

\subsection{Baselines for Identifying Vulnerable Records}
\label{subsec:baselines}
We compare our approach against three baselines to identify vulnerable records:

\textbf{Random}: randomly sample target records from the entire population.

\textbf{Rare Value (Groundhog)}~\cite{stadler2022synthetic} sample a target record that either has a rare value for a categorical attribute or a value for a continuous attribute that is larger than the 95th percentile of the respective attribute.

\textbf{Log-likelihood (TAPAS)}~\cite{houssiau2022tapas}: sample target records having the lowest log-likelihood, assuming attribute values are independently drawn from their respective empirical distributions.

Note that this approach is defined by the authors only for categorical attributes $f \in \mathcal{F}_{\text{cat}}$: for each record $x_i$, they compute the frequency of the value $x_{i,f}$ in the entire dataset, resulting in $p_{i,f}$. 
The likelihood of a record is defined by the product of $p_{j,f}$ across all attributes. 
We here extend this approach to also account for continuous attributes $f\in\mathcal{F}_{\text{cont}}$ by discretising them according to percentiles.

To evaluate the different methods to select vulnerable records, we run the query-based MIA described in Sec.~\ref{subsec:SOTA_MIA} on the ten records identified by each of the three methods as well as our own approach. 

\subsection{Synthetic Data Generation Models}
\label{subsec:generators}
We evaluate our results against three generative models using the implementation available in the \texttt{reprosyn} repository \cite{reprosyn2022}. 

\textbf{SynthPop} first estimates the joint distribution of the original, private data to then generate synthetic records. This joint distribution consists of a series of conditional probabilities which are fitted using classification and regression trees. With a first attribute randomly selected, the distribution of each other attribute is sequentially estimated based on both observed variables and previously generated synthetic columns. Initially proposed as an R package \cite{nowok2016synthpop}, the reprosyn repository uses a re-implementation in Python~\cite{synthpop2022}.

\begin{table}[htbp!]
    \centering
    \begin{tabular}{|c|c|c|c|c|}
    \hline
         &|$D_{aux}$|& |$D_{test}$| & $n_{shadow}$ & $n_{test}$\\
         \hline
        Adult & 10000 & 5000 & 4000 & 200\\
        \hline
        UK Census & 50000 & 25000 & 4000 & 200\\
        \hline
    \end{tabular}
    \caption{Different parameters used throughout experiments}
    \label{tab:parameters}
\end{table}

\textbf{BayNet} uses Bayesian Networks to model the causal relationships between attributes in a given dataset. Specifically, the attributes are represented as a Directed Acyclic Graph, where each node is associated with an attribute and the directed edges model the conditional independence between the corresponding attributes. The corresponding conditional distribution of probabilities $\mathbb{P}[X|Parents(X)]$ is estimated on the data using a GreedyBayes algorithm (for more details we refer to Zhang et al. \cite{zhang2017}). Synthetic records can then be sampled from the product of the computed conditionals, i.e. the joint distribution.

\textbf{PrivBayes}
\label{subsec:privBayes} is a differentially private version of the BayNet algorithm described above. Here, a first Bayesian Network is trained under an $\epsilon_1$-DP algorithm. The conditional probabilities are then computed using an $\epsilon_2$-DP technique, leading to a final $\epsilon = (\epsilon_1 + \epsilon_2)$-DP mechanism. From this approximate distribution, we can generate synthetic records without any additional cost of privacy budget. Note that for $\epsilon \to \infty$ PrivBayes becomes equivalent to BayNet. Again, we refer the reader to Zhang et al. \cite{zhang2017} for a more in-depth description.

\subsection{Datasets}
\label{subsec:datasets}
We evaluate our method against two publicly available datasets. 

\textbf{UK Census~\cite{census2011}} is the 2011 Census Microdata Teaching File published by the UK Office for National Statistics. 
It contains an anonymised fraction of the 2011 Census from England and Wales File (1\%), consisting of 569741 records with 17 categorical columns. 

\textbf{Adult~\cite{adult1996}} contains an anonymized set of records from the 1994 US Census database. 
It consists of 48842 records of 15 columns. 
9 of those columns are categorical, and 6 are continuous. 

\subsection{Parameters of the Attack}
\label{subsec:params}

Table \ref{tab:parameters} shows the parameters we use for the attack. We use a given dataset $\Omega$ (e.g. Adult) to select our most at-risk records as the target records.
We then partition $\Omega$ randomly into $D_{test}$ and $D_{aux}$.   
$D_{aux}$ is the auxiliary knowledge made available to the attacker, while $D_{test}$ is used to sample $n_{test}$ datasets used to compute the attack performance.
$n_{shadow}$ represents the number of shadow datasets used, sampled from the auxiliary knowledge $D_{aux}$. 
We considered, in all experiments, the size of the release synthetic dataset ($D^s$) to be equal to the size of the private dataset $D$, $|D|=|D^s|=1000$.
To create the shadow datasets, while ensuring that $D_{aux}$ does not contain the target record, we randomly sample $|D|-1$ records in $D_{aux}$. Then we add the target record $x_T$ in half of the datasets, and in the other half, we used another record randomly sampled from $D_{aux}$.

We run the query-based attack using $N=100000$ queries randomly sampled from all possible count queries ($2^F$ in total) and a random forest meta-classifier with 100 trees, each with a maximum depth of 10.

In order to measure the attack performance, we compute the Area Under the receiver operating characteristic Curve (AUC) for the binary classification of membership on $n_{test}$ datasets.

Lastly, in line with the auditing scenario, we use $k=5$ neighbors in the private dataset $\Omega$ to select the most vulnerable records using our vulnerability score $V_k$. 

\begin{figure}[!ht]
\centering
\includegraphics[width=0.49\linewidth]{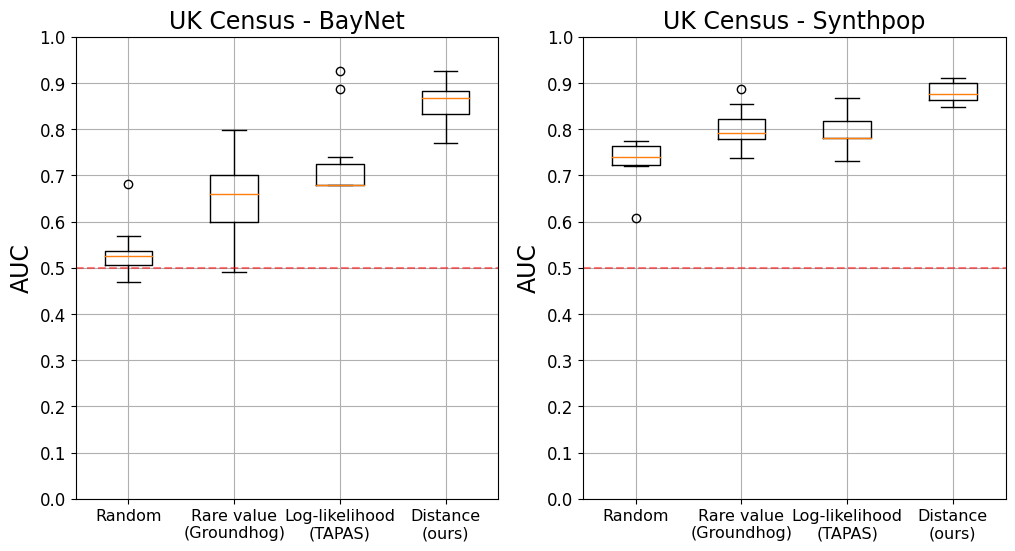} 
\includegraphics[width=0.49\linewidth]{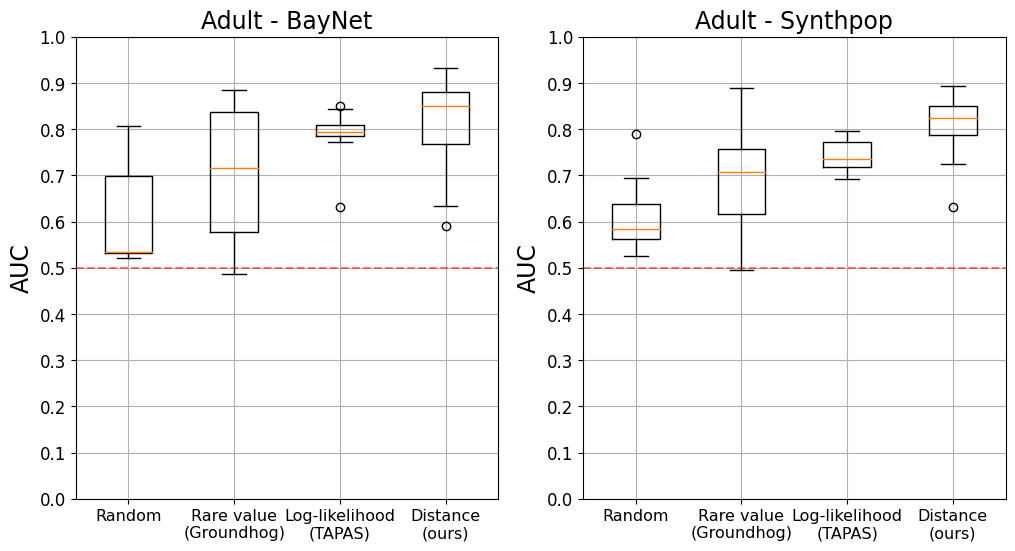} 
    \caption{AUC of MIAs across record identification techniques for the query-based attack, for datasets UK Census and Adult and synthetic data generators BayNet and Synthpop.} 
\label{fig:query_primary_results}
\end{figure}

\begin{table}[!ht]
  \centering
  \caption{Mean and standard deviation for AUC for the query-based attack method}
  \label{tab:meanstd_query_results}
  \begin{tabular}{|l|c|c|c|c|}
    \hline
    \multicolumn{1}{|c|}{\multirow{2}{*}{\textbf{Method}}} & \multicolumn{2}{c|}{\textbf{UK census}}&  \multicolumn{2}{c|}{\textbf{Adult}}\\
    &\multicolumn{1}{c}{Synthpop}&BayNet&\multicolumn{1}{c}{Synthpop}& BayNet\\
    \hline
    Random & $0.732 \pm 0.046$ & $0.535 \pm 0.055$ & $0.613 \pm 0.084$ & $0.618 \pm 0.120$ \\
    \hline
    Rare value (Groundhog) & $0.802 \pm 0.044$ & $0.644 \pm 0.086$ & $0.699 \pm 0.113$ & $0.704 \pm 0.140$\\
    \hline
    Log-likelihood (TAPAS) & $0.790 \pm 0.041$ & $0.731 \pm 0.090$ & $0.742 \pm 0.033$ & $0.787 \pm 0.057$\\
    \hline
    Distance (ours) & \bm{$0.879 \pm 0.021$} & \bm{$0.858 \pm 0.040$} & \bm{$0.804 \pm 0.072$} & \bm{$0.810 \pm 0.110$}\\
    \hline
  \end{tabular}
\end{table}

\section{Results}
\label{sec:results}
\subsection{Performance of Vulnerable Record Identification Methods}

We evaluate the effectiveness of vulnerable record identification methods across the two datasets UK Census and Adult, and two synthetic data generators Synthpop and BayNet. For each dataset, each of the four record identification methods selects 10 target records. We then perform the MIA, as a privacy game, on each of these records and report its AUC.

Figure \ref{fig:query_primary_results} show that our vulnerable record identification method consistently outperforms both the random baseline and the two ad-hoc techniques previously used in the literature. The median risk of records identified by our method is indeed consistently higher than all other methods, up to 18.6 p.p. higher on the UK Census dataset using BayNet. On average, the AUC of the records we identified are 7.2 p.p. higher than the AUC of the records selected by other methods (Table \ref{tab:meanstd_query_results}). Importantly, our method also consistently manages to identify records that are more or equivalently at-risk compared to other methods.

\subsection{Applicability of the Method to New Attacks}

MIAs against synthetic data is an active area of research with a competition \cite{snake_challenge} now being organised to develop better, more accurate, MIAs. To evaluate the effectiveness of our vulnerable record identification, we developed a new attack which we aimed to be as different as possible to the existing state of the art. We call this new attack \emph{target attention}.

\begin{figure}[!ht]
\centering
\includegraphics[width=0.49\linewidth]{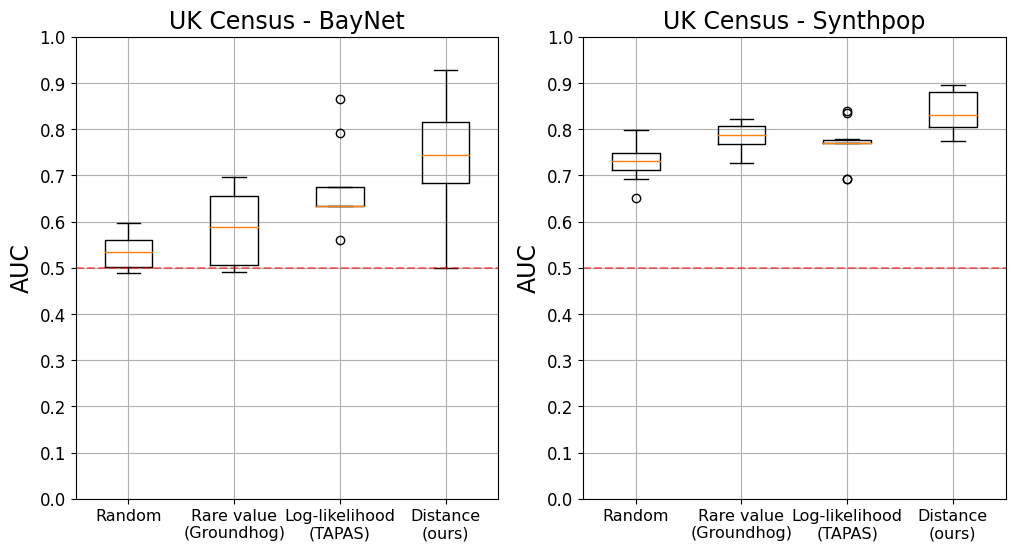} 
\includegraphics[width=0.49\linewidth]{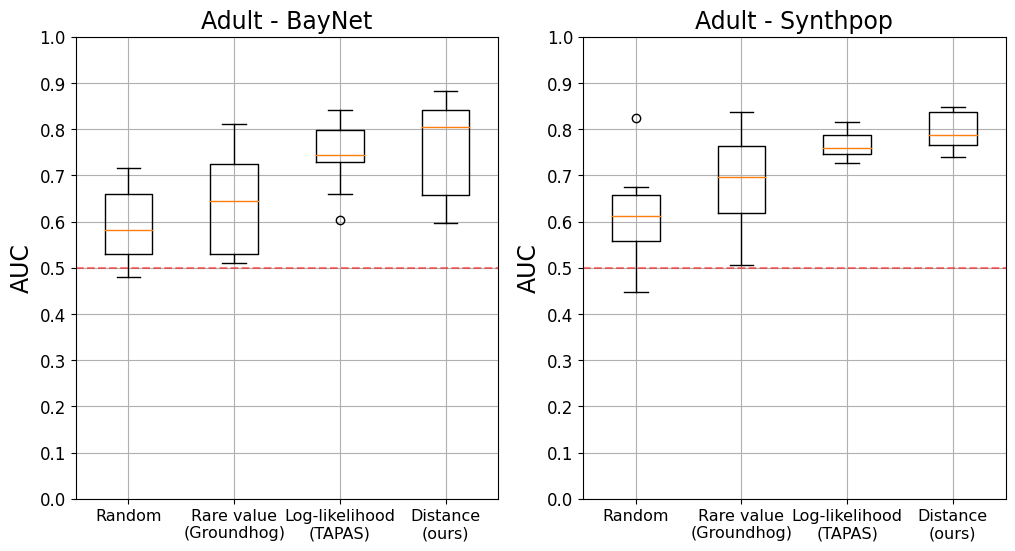} 
    \caption{AUC of MIAs across record identification techniques for the target attention attack, for datasets UK Census and Adult and synthetic data generators BayNet and Synthpop.} 
\label{fig:targetattention_primary_results}
\end{figure}

In contrast with the state-of-the-art method, which first computes features from a synthetic dataset to then use these as input for a meta-classifier, our new attack takes as input (a part of) the synthetic dataset to directly predict membership. As such, the model would be able to extract directly from the synthetic dataset the information that is most useful for the MIA. To our knowledge, it is the first method that is record-focused and allows for trainable feature extraction. For more information about the new method, we refer the reader to Appendix \ref{app:targetattention}. We then follow the same methodology as above to evaluate the effectiveness of the four vulnerable record identification methods when using the target attention approach.

Figure \ref{fig:targetattention_primary_results} shows that, again, our vulnerable record identification method strongly outperforms both the random baseline and the two ad-hoc techniques previously used in the literature. The mean risk of records identified by our method is again consistently higher than all other methods, up to 10.89 p.p. higher on UK Census using BayNet. Across datasets and generators, the records we identify are 5.2 p.p. more at risk than the records selected by other methods (Appendix \ref{app:targetattention}).

\subsection{Robustness Analysis for $k$}

Throughout the paper, we use $k=5$, meaning that our vulnerability score $V_k(x_i)$ is computed using the 5 closest records to the record of interest. We here evaluate the impact of a specific choice of $k$ for the record identification and associated results. More specifically, we compute the mean distance for each record using values of $k$ ranging from 1 to 50 and report the mean AUC for the ten records selected by the method on the Adult dataset using Synthpop. 

\begin{figure}[!ht]
\centering
\subfigure[]{
\includegraphics[width=0.4\linewidth]{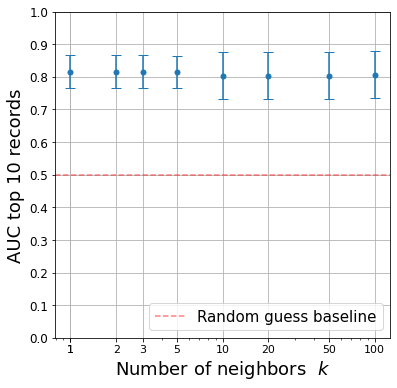}}
\subfigure[]{
\includegraphics[width=0.4 \linewidth]{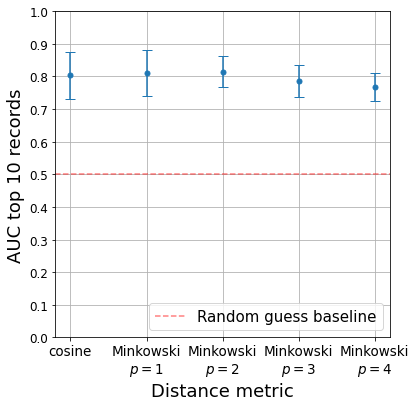}
}
    \caption{Mean and standard deviation of the AUC of MIAs for 10 records selected by the vulnerability score for varying (a) number of neighbors considered and (b) distance metrics considered. Results on Adult, Synthpop for the query-based attack.} 
\label{fig:determining_k_metric}
\end{figure}

Figure \ref{fig:determining_k_metric}(a) shows our method to be robust to specific choices of $k$. The mean AUC of the top ten records varies indeed only slightly for different values of $k$. Looking at the top 10 records selected for different values of $k$, we find that only a handful of records are distinct, suggesting our method to not be too sensitive to specific values of $k$.

\subsection{Robustness Analysis for the Cosine Distance}

We here rely on a generalized form of cosine distance across attribute types. We now evaluate the susceptibility of our results to different choices of distance metrics. We do this using several versions of the Minkowski distance between two vectors $a, b \in \mathbb
{R}^L$. For a given value $p \in \mathbb{N}^*$, the Minkowski distance is defined as $d_{\text{Minkowski}}^p(a, b) = \sqrt[p]{\sum_{l=1}^{L} |a_l - b_l|^p}$.
Similarly to Eq. \ref{eq:distance}, we generalise this distance metric for two records $x_i, x_j\in D$ to:\begin{equation}d(x_i, x_j) := \frac{|\mathcal{F}_{\text{cat}}|}{F}     d_{\text{Minkowski}}^p(h(x_i) , h(x_j)) 
    + \frac{|\mathcal{F}_{\text{cont}}|}{F} d_{\text{Minkowski}}^p(c(x_i) , c(x_j)).
\label{eq:distance_minkowski}
\end{equation}

Notably, the Minkowski distance with $p=2$ closely relates to the cosine distance. As before, we compute the mean distance for each record using Minkowski distances with $p$ ranging from 1 to 4 and report the mean AUC for the ten records selected by the method on the Adult dataset using Synthpop.

Figure \ref{fig:determining_k_metric}(b) shows the attack performance to, again, be robust to different choices of distance metric. In particular, mean AUC is not significantly different for the cosine distance and $p\in \{1,2\}$, while slightly decreasing for higher $p$ values. Cosine being the standard distance metric in the literature, we prefer to use it.

\subsection{Vulnerable Records and Differentially Private Generators }
\label{subsec:DP}

An active line of research aims at developing differentially private (DP) synthetic data generators. As discussed above however, developing models that achieve DP with meaningful theoretical $\epsilon$ while maintaining high utility across tasks remains a challenge. MIAs are thus an important tool to understand the privacy protection offered by DP models for various values of $\epsilon$.

\begin{figure}[!ht]
\centering
\includegraphics[width=0.4\linewidth]{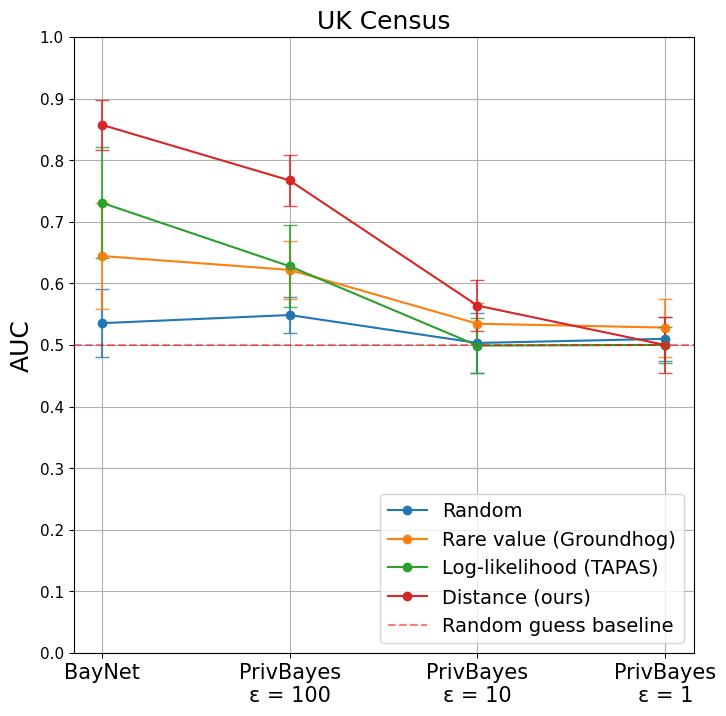} 
\includegraphics[width=0.4\linewidth]{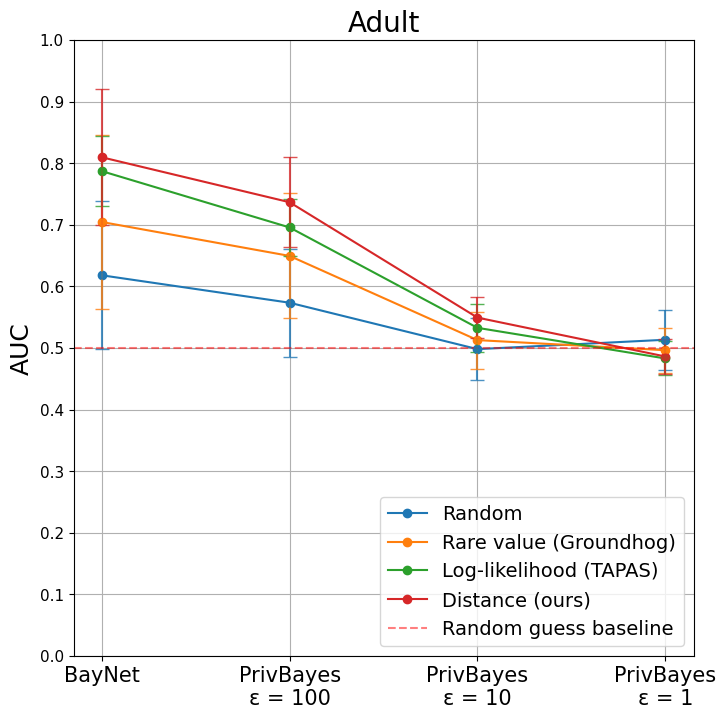} 
    \caption{Mean and standard deviation of the AUC of the MIAs for varying values of $\epsilon$ in DP-based generator PrivBayes. Results on UK Census (left) and Adult (right) for the query-based attack and for the 10 points for each record identification method.} 
\label{fig:DP_results}
\end{figure}

We here evaluate the performance of our vulnerable record identification approach when DP generators are used. Specifically, we report the mean and standard deviation AUC for the 10 records identified as being at risk by our approach and by the three baselines for varying values of $\epsilon$ using PrivBayes \cite{zhang2017}. 

Figure \ref{fig:DP_results} shows that our approach outperforms both the random baseline and the two ad-hoc techniques previously used in the literature. As expected though, the AUC drops when $\epsilon$ decreases. The attack still performs quite well for $\epsilon=100$ and manages to identify at-risk records for $\epsilon=10$. Importantly, the general utility of synthetic data generated by PrivBayes for low value of $\epsilon$ is being debated.

\section{Discussion}
\label{sec:discussion}

Our results show that our simple distance-based method is able to consistently identify vulnerable records, and more effectively so than previously proposed ad-hoc methods. We believe this to be due to limitations of previous work. Indeed \emph{Rare value (Groundhog)} only considers the value of one specific attribute to determine whether a record is at risk while \emph{Log-likelihood (TAPAS)} assumes independence across all attributes to compute a record's likelihood, an assumption that is typically not met in real-world datasets. In contrast, our distance-based metric considers the neighborhood of the record, therefore encompassing all attributes without making any assumptions on the underlying distribution. 

Across datasets and generators, we find that records identified by our method are 
on average 7.2 p.p. more vulnerable than previous work when using the state-of-the-art MIA. In practice, we believe this to be highly significant. Researchers and practitioners who wish to evaluate new attack methods and the privacy risk of synthetic data publishing are indeed unable to evaluate all records in the dataset, due to the computational cost of shadow modeling. Accurately identifying the records that are most at risk is thus as essential as using the state-of-the-art MIA when evaluating the privacy risk of a synthetic dataset. Indeed, data protection laws such as the EU GDPR require to consider the privacy protection offered to all data subjects when evaluating if a data release is anonymous.

 When we use only categorical columns, such as in UK Census, results show that we reduce the standard deviation in the results by a factor of two, across generators. With our method we obtained a standard deviation of 0.021 instead of 0.041 for Synthpop and 0.040 instead of 0.090 for BayNet. This shows that our method selects records with high precision of having high vulnerability. 

Lastly, apart from being effective, our principled method is compellingly easy. By being independent of the synthetic data generator and the dataset, it allows for pragmatic and efficient cross-generator and dataset comparison.  

\textbf{Future Work.} We here distinguished between categorical features and continuous attributes. 
Future research could explore if considering ordinal as a subset of the categorical features could lead to a better record selection. Similarly, we here focused on the difference in metrics between the cosine similarity and the Minkowski metrics of order $p \in \{1,2,3,4\}$. We leave for future work to see if other distances, such as Minkowski of higher order, the Chebyshev distance or Hamming distance, could lead to a better record selection. 

Additionally, one could study whether removing the most vulnerable records is an effective defense against attacks. Prior work on MIAs against ML models suggests that, when vulnerable records are removed, other records then become more vulnerable \cite{carlini2022privacy}. We leave for future work whether this holds true for MIAs against synthetic data. 

Finally, we presented results on three widely used synthetic data generators. Future work should evaluate whether more specific metrics could be beneficial for other generators such as GANs.  

\section{Conclusion}
\label{sec:conclusion}

Due to computational costs, the identification of vulnerable records is as essential as the actual MIA methodology when evaluating the privacy risks of synthetic data, including with respect to data protection regulations such as the EU GDPR. We here propose and extensively evaluate a simple yet effective method to select vulnerable records in synthetic data publishing, which we show to strongly outperform previous ad-hoc approaches.

\subsubsection{Acknowledgements} 
We acknowledge computational resources and support provided by the Imperial College Research Computing Service. \footnote{\url{http://doi.org/10.14469/hpc/2232}.}

\bibliographystyle{splncs04}
\bibliography{mybibliography}

\appendix
\section{Appendix: Target Attention}
\label{app:targetattention}

Inspired by developments in natural language processing \cite{vaswani2017attention}, the target attention model computes record-level embeddings. Through a modified attention mechanism, these embedding are able to interact with the embedding of the target record, which in turn leads to a dataset-level embedding that is used as input for a binary classifier. The detailed steps are laid out below. 

We start by preprocessing the synthetic dataset used as input $D^s$. First, we compute the one-hot-encoded values for all categorical attributes and apply this consistently to all the synthetic records and the target record. The continuous attributes remain untouched.
Second, we compute all the unique records in the dataset and their multiplicity. 
Third, we compute the distance following equation (1) between all unique synthetic records and the target record.
Finally, we rank all unique synthetic record by increasing distance to the target record and, optionally, only keep the \emph{top $X$} closest records, where $X$ is a parameter.

The final pre-processed synthetic dataset $D^{s, p} \in \mathbb{R}^{X \times F'}$ is a matrix containing the top \emph{X} unique records, their multiplicity and the distance, ranked by the latter, where $F'$ denotes the number of features after one-hot-encoding of the categorical attributes added by 2 to account for the multiplicity and the distance. For consistency, the target record $x_t$ is processed in the same way, with one-hot-encoded categorical attributes, multiplicity $1$ and distance $0$, resulting in the pre-processed target record $x_{t, p} \in \mathbb{R}^{F'}$.

\begin{figure}[!ht]
\centering
\includegraphics[width=0.7\linewidth]{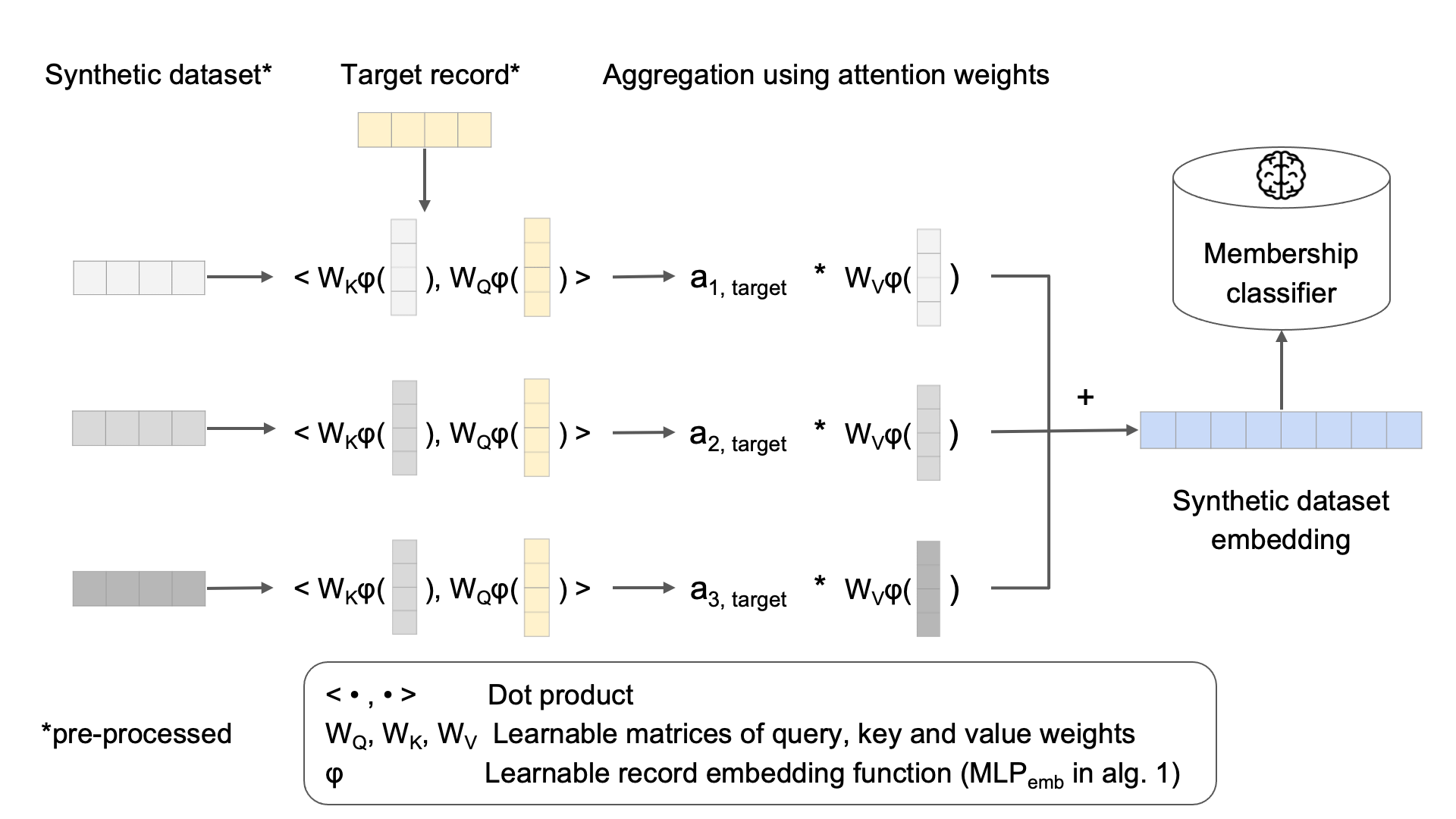} 
    \caption{Illustration of our target attention architecture.} 
\label{fig:target_attention}
\end{figure}

Fig.~\ref{fig:target_attention} illustrates our target attention model, which we describe in Alg.~\ref{alg:target_attention}.
The model takes as input the pre-processed synthetic dataset $D^{s, p}$. 
First, a multilayer perceptron (MLP) is used to compute embeddings of both the pre-processed records as well as the target record. Second, after computing a query vector for the target record and keys and values for the records, an attention score is computed for each record. After applying a \emph{softmax} function across all attention scores, all record values are summed weighted by the respective attention score to eventually lead to the dataset-level embedding which is used as input for a final MLP as binary classifier. The output of the entire network is a single float, which after applying a \emph{sigmoid} function $\sigma$ results in a probability of membership. This model is then trained with all synthetic shadow datasets using a binary cross-entropy loss function with the known binary value for membership. 

\begin{algorithm}[!ht]
  \caption{Forward propagation for target-attention}
  \label{alg:target_attention}
  \begin{algorithmic}[1]
    \Inputs{1. $D^{s, p} \in \mathbb{R}^{X \times F'}$ the preprocessed synthetic dataset with its unique records, their multiplicity and distance to the target record. \\
    2. The pre-processed target record values $x_{t, p} \in \mathbb{R}^{F'}$. \\
    3. Learnable MLP for the embedding $\textbf{MLP}_{emb}$ with $F'$ input features and 
    $F_{emb}$ output features, weight matrices $\textbf{W}_q, \textbf{W}_k$, and $\textbf{W}_v \in \mathbb{R}^{F_{emb} \times F_{att}}$ to compute the target query and record keys and values with attention size $F_{att}$, and MLP for the prediction $\textbf{MLP}_{pred}$ with $F_{att}$ input features and $1$ output feature}
    \Output{${h}_{pred} \in \mathbb{R}$, which leads to the predicted probability for binary membership after applying a sigmoid transformation $\sigma({h}_{pred})$}
    \State{\comm{Compute the embeddings of all records and the target record}} \\
    $D_{emb} \gets \textbf{MLP}_{emb}[D^{s, p}]$ with $D_{emb} \in \mathbb{R}^{X \times F_{emb}}$ \\
    $x_{t, emb} \gets \textbf{MLP}_{emb}[x_{t, p}]$ with $x_{t, emb} \in \mathbb{R}^{1 \times F_{emb}}$
    \State{\comm{Compute the target query, and the keys and values for each record}} \\
    $q_t \gets \textbf{W}_q x_{t, emb}$ with $q_t \in \mathbb{R}^{1 \times F_{att}}$ \\
    $K \gets \textbf{W}_k D_{emb}$ with $K \in \mathbb{R}^{X \times F_{att}}$ \\
    $V \gets \textbf{W}_v D_{emb}$ with $V \in \mathbb{R}^{X \times F_{att}}$
    \State{\comm{Compute the target attention scores for each record}} \\
    $a_j \gets <q_t, v_j>$ for $j=1..X$ to get $a \in \mathbb{R}^{1 \times X}$ \\
    $a_j \gets \exp{a_j} / (\sum_{j=1}^{X} \exp{a_j)}$ for $j=1..X$ \\
    $D_{att} \gets a V $ with $D_{att} \in \mathbb{R}^{1 \times F_{att}}$ \comm{Sum values weighted by target attention} \\
    $\textbf{h}_{pred} \gets \sigma(\textbf{MLP}_{pred}[D_{att}])$ \comm{Compute the final membership score.}
  \end{algorithmic}
\end{algorithm}

Throughout the experiments, we used $F_{emb} = 20$, $ F_{att} = 15$ and $|top~X| = 100$. Both $MLP_{emb}$ and $MLP_{pred}$ consist of one hidden layer with 20 and 10 nodes respectively, and use ReLU as activation and a dropout ratio of 0.15. The network is trained for 500 epochs, using the Adamax optimizer \cite{kingma2014adam} with a learning rate of $10e-3$ and batch size of $20$. 10\% of the training data is used as validation data and the model that achieves the lowest validation loss is selected. Results in Table \ref{tab:meanstd_target_attention_results} display the details of results compiled in Figure \ref{fig:targetattention_primary_results}. 

\begin{table}[!ht]
  \centering
  \caption{Mean and standard deviation for AUC for the Target Attention method}
  \label{tab:meanstd_target_attention_results}
  \begin{tabular}{|l|c|c|c|c|}
    \hline
    \multicolumn{1}{|c|}{\multirow{2}{*}{\textbf{Method}}} & \multicolumn{2}{c|}{\textbf{UK census}}&  \multicolumn{2}{c|}{\textbf{Adult}}\\
    &\multicolumn{1}{c}{Synthpop} & BayNet & \multicolumn{1}{c}{Synthpop} & BayNet\\
    \hline
    Random & $0.727 \pm 0.038$ & $0.535 \pm 0.035$ & $0.609 \pm 0.112$ & $0.594 \pm 0.083$ \\
    \hline
    Rare value (Groundhog) & $0.783 \pm 0.030$ & $0.587 \pm 0.079$ & $0.691 \pm 0.103$ & $0.638 \pm 0.104$\\
    \hline
    Log-likelihood (TAPAS) & $0.770 \pm 0.046$ & $0.674 \pm 0.085$ & $0.765 \pm 0.027$ & $0.745 \pm 0.068$\\
    \hline
    Distance (ours) & \bm{$0.838 \pm 0.041$} & \bm{$0.737 \pm 0.118$} & \bm{$0.797 \pm 0.038$} & \bm{$0.763 \pm 0.103$}\\
    \hline
  \end{tabular}
\end{table} 

\end{document}